\documentclass[twocolumn,aps,showpacs,preprintnumbers,superscriptaddress]{revtex4}
\usepackage{graphicx}
\usepackage{bm}
\usepackage{amsmath, amsthm, amssymb}

\begin{document}
\title{Generalized Parametric Space, Parity Symmetry of Reflection, and Systematic Design Approach for Parity-Time Symmetric Photonic Systems}
\author{Jeng Yi Lee}
\affiliation{
Department of Opto-Electronic Engineering, National Dong Hwa University, Hualien 974301, Taiwan }
\author{Pai-Yen Chen}
\affiliation{Department of Electrical and Computer Engineering, University of Illinois at Chicago, Chicago, IL 60661,
USA}

\date{\today}

\begin{abstract} 
Based on the reciprocity theorem, we put forward a generalized parametric space for arbitrary transfer matrix with parity time (PT) symmetry. 
Through this space, one can extract complete information involving PT phases, reflectances, transmittance and known extraordinary scattering phenomena. 
We demonstrate a PT heterostructure with coherent perfect absorption-lasing, anisotropic transmission resonance, and parity symmetry of reflection coefficients at the frequencies of interest.
In addition, with the parametric space and the analytical formula, the corresponding complex dielectric permittivities for a simple PT system made of a gain, a gap, and a loss media in deeply subwavelength is derived to achieve various exotic PT functionality. 
This work could offer an alternative route to design versatile optical and photonic PT devices.
\end{abstract}
\pacs{ }

\maketitle

\section{Introduction}

In photonics, a system with parity time (PT) symmetry could be implemented by a delicate balance of gain (amplifying) and loss (attenuation) media in spatial placement, i.e., $\epsilon(\vec{r})=\epsilon^{*}(-\vec{r})$ where $\epsilon$ is complex dielectric permittivity.
As light transversely propagates through one dimensional PT heterostructures, the scattering response can be classified into two phase: symmetry and broken symmetry \cite{yidong1}.

In the symmetry phase, the corresponding eigenvalues of scattering matrix are both unimodular and formed complex conjugate pairs \cite{yidong1,yidong2,annphys,invisible1}.
In the broken symmetry phase, the eigenvalues would form reciprocal pairs and nonunimodular, corresponding to amplifying and attenuation.
In particular, a PT system can behave a coherent perfect absorption and lasing (CPAL) at the same time, with the corresponding zero and pole eigenvalues coincided\cite{yidong1,cpal}.
CPAL systems can be switchable by its proper input waves, with practical applications in highly sensitive sensors \cite{yen1}.
This CPAL was experimentally demonstrated in Refs. \cite{ptlaser,ptlaser1}.
In principle, reflectances of PT systems are anisotropic. 
This anisotropic property would lead to asymmetrical forces by opposite illuminations  \cite{force1, force2}.
Between the symmetry and broken symmetry, there has the exceptional point, associated with degenerate eigenstates, that it allows for observation of unidirectional invisibility \cite{invisible1, invisible2}, that could not occur at any unitary systems.
On the other hand, bi-reflectionless and a non-zero transmission phase are observed in the work \cite{yidong2}.
Numerous functional devices incorporating with PT symmetry have been observed, including  power oscillation \cite{doubleref}, Bloch oscillation \cite{bloch}, and negative refraction \cite{negref}.

Although a balance of gain and loss in space is a fingerprint for PT symmetry, the CPAL result can be achieved in unbalance gain and loss \cite{yen2,yen3}.
Even system is made of only passive media, anisotropic reflectionless had been experimentally observed, associated with emergence of an exceptional point\cite{ol}.
In addition, a large class of non-Hermitian systems without PT symmetry can support exceptional points, leading to applications in enhanced sensing \cite{general1,Li1}.
Discovered from extraordinary wave manipulation, PT and non-Hermitian systems have became attractive studies in a variety of wave subjects \cite{general1,Li1,general2,review1,acoustic1,acoustic2,acoustic3,elastic1}.

In this work, we propose a generalized parametric space for arbitrary PT transfer matrix by exploiting its intrinsic symmetry and reciprocity theorem.
With the systematic diagram proposed here, we indicate extraordinary scattering phenomena, reflectances, and transmittance, irrespective of inherent system configuration.
As representative examples, we discuss a simple PT system by sweeping frequency to observe CPAL, anisotropic transmission resonance (ATR), and parity symmetry of reflection coefficients (PSR).
Furthermore, we derive permittivity expressions for another simple PT system made by a gain, a gap, a loss to manipulate wave response.
This work could offer an effective route to design PT devices.

\begin{figure*}
\centering
\includegraphics[width=16cm]{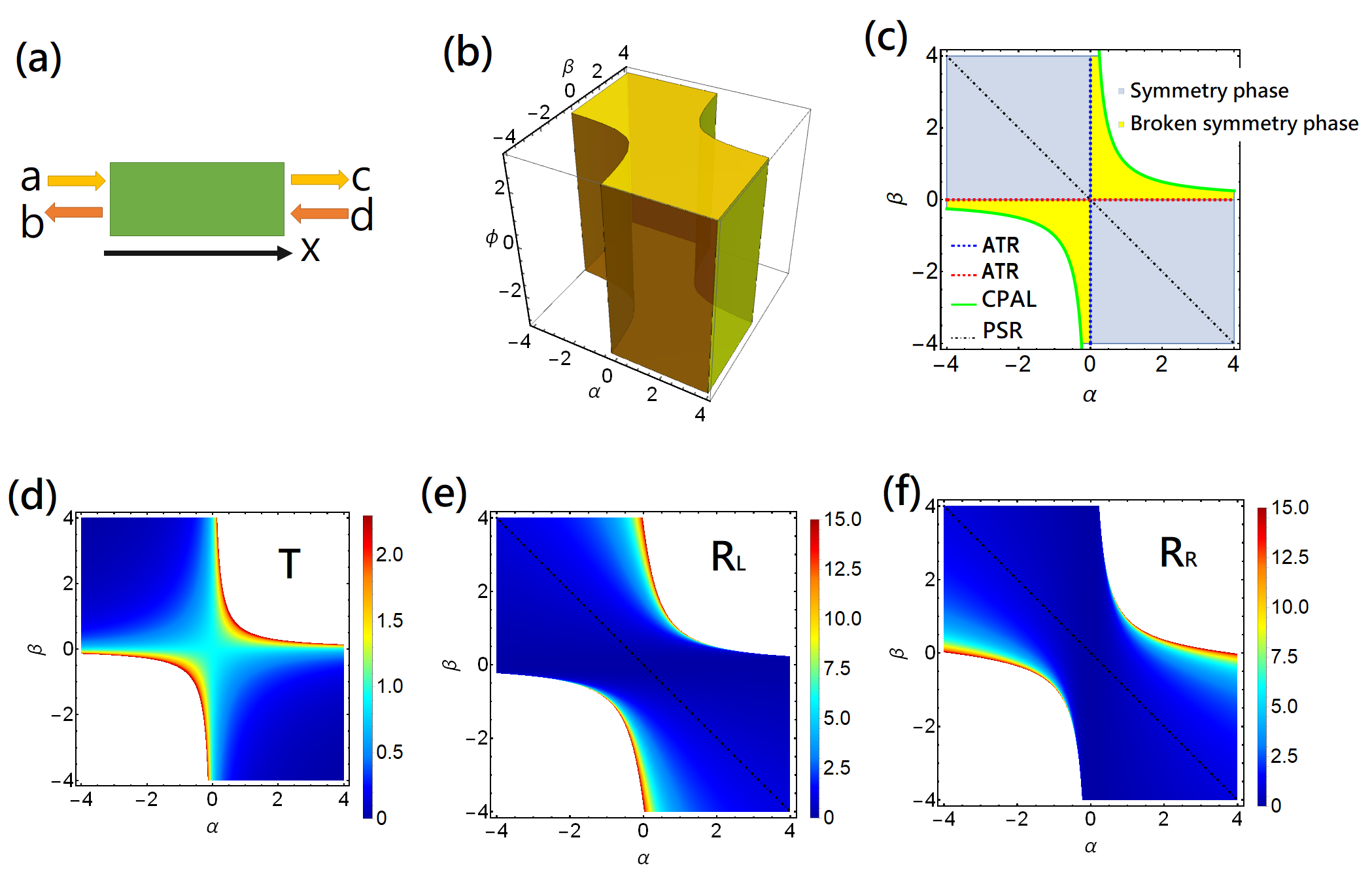}
\caption{(Color online) (a) Schematic configuration of one-dimensional scattering system interacted with two independent input waves $a$ and $d$ and accompanied with output waves $b$ and $c$.  (b) Three-dimensional parametric space in terms of $\alpha$, $\beta$, and $\phi$ for general PT transfer matrix.
Allowed parametric space is depicted by yellow. In our context, we find extraordinary PT scattering phenomena are related to parameters $\alpha$ and $\beta$. We thus plot 2D parametric diagram with $\alpha$ and $\beta$ in which allowed parameters are illustrated by gray and bright yellow colors. The gray region denotes PT symmetry phases, while the bright yellow denotes broken symmetry  phase. Between these two phases, there has an exceptional point regime associated with degeneracy of eigenvalues, marked by blue and red dashed lines, corresponding to anisotropic  transmission resonance (ATR). CPAL stands for coherent perfect absorber-laser, corresponding to green line, located at the allowed-forbidden boundary. Based on this 2D parametric diagram, we plot transmittance T, reflectances for left and right incidences, $R_L$ and $R_R$, respectively.} 
\end{figure*} 

\section{Parametric space for arbitrary PT systems}
We consider that  a one-dimensional PT system in a homogeneous environment interacts with two independent input waves, as shown in Fig. 1 (a).
 A set of linear relation for total waves in the left and right leads can be expressed as \cite{book1}
\begin{equation}
\begin{bmatrix}
c\\
d\\
\end{bmatrix}=
M
\begin{bmatrix}
a\\
b\\
\end{bmatrix}
=\begin{bmatrix}
M_{11} & M_{12}\\
M_{21} & M_{22}\\
\end{bmatrix}
\begin{bmatrix}
a\\
b\\
\end{bmatrix}
\end{equation}
here $M$ is transfer matrix. $ae^{ik_0x}$ ($ce^{ik_0x}$) and $be^{-ik_0x}$ ($de^{ik_0x}$) are right and left travelling waves, respectively, with $a$ ($c$) and $b$ ($d$) being complex amplitudes. $k_0$ is wavenumber outside the system.
When a system has PT-symmetry invariant, the transfer matrix obeys the following property: $M^{-1}=M^{*}$ \cite{cpal,yidong1}. 
Thus $M_{11}^{*}=M_{22}$ and $M_{12(21)}^{*}=-M_{12(21)}$. 
Combined with fundamental optical reciprocity,  det$M=1$ is always satisfied \cite{book2}.
Thereby, we can assign
\begin{equation}
\begin{split}
M_{12} &=i\alpha
,M_{21} =i\beta\\
\vert M_{11}\vert^2 &=1-\alpha\beta \geq 0\\
\rightarrow M_{11} &=\sqrt{1-\alpha\beta}e^{i\phi}
\end{split}
\end{equation}
here $\alpha$ and $\beta$ are two real numbers and $\phi$ is transmission phase.
We remark that even system violates PT symmetry, optical reciprocity can still be met\cite{Li1,book2}.

Therefore,  the complete description of arbitrary PT transfer matrix would rely on $\alpha$, $\beta$, and $\phi$.
We note that similar outcomes have already proposed in Refs \cite{yidong2}.
However, we can further display the parametrization by an visual 3D diagram, depicted in Fig. 1 (b).
Within the yellow color, it presents the allowed parameters for any PT transfer matrix, i.e., $\alpha\beta \leq 1$ and $\phi \in [-\pi,\pi]$.

The scattering matrix $S$ \cite{note}, connecting out-going waves with incoming wave, is
\begin{equation}
\begin{split}
\begin{bmatrix}
c\\
b
\end{bmatrix}=S
\begin{bmatrix}
a\\
d
\end{bmatrix}=
\begin{bmatrix}
t & r_R\\
r_L & t
\end{bmatrix}
\begin{bmatrix}
a\\
d
\end{bmatrix}.
\end{split}
\end{equation}
By the parametrization, the transmission coefficient is $t=\frac{e^{i\phi}}{\sqrt{1-\alpha\beta}}$, while reflection coefficients for left incidence $r_L$ and for right incidence $r_R$ are $r_L=\frac{\beta e^{i(\phi-\frac{\pi}{2})}}{\sqrt{1-\alpha\beta}}$ and $r_R=\frac{\alpha e^{i(\phi+\frac{\pi}{2})}}{\sqrt{1-\alpha\beta}}$, respectively.
The corresponding eigenvalues are $S_{\pm}=\frac{e^{i\phi}}{\sqrt{1-\alpha\beta}}(1\pm \sqrt{\alpha\beta})$ while eigenvectors are $\vert S_{\pm}>=[\pm i\frac{\sqrt{\alpha}}{\beta},1]^{T}$ .

For PT-symmetry systems, a hybrid functionality of coherent perfect absorption and lasing (CPAL) can be simultaneously existed \cite{yidong1,cpal}, when $M_{11}=M^{*}_{22}=0$ can be satisfied.
Despite that, the final scattering response is still determined by two input waves.
To perform coherent perfect absorption, i.e., all input waves are fully absorbed without output waves, it restricts input waves $b=M_{21}a=i\beta a$.
Deviation from this condition, PT system would exhibit lasing.
This small perturbation of input waves resulting in  drastic changes of output waves  has practical potentials in design of detectors \cite{yen1}.
In parametrization, the operation of this hybrid functionality needs $\alpha\beta=1$, independent of transmission phase $\phi$.
Then, the corresponding transmittance ($T=\vert t\vert^2$) and two reflectances ($R_R=\vert r_R\vert^2$ and $R_L=\vert r_L\vert^2$) would be all infinity under only single beam illumination.

PT system can also exhibit anisotropic transmission resonance (ATR).
It can have reflectionless from one side illumination and unity transmittance, but would 
 encounter enhanced reflection from another side illumination \cite{invisible1,invisible2}, that can not occur in any unitary systems.
ATR needs $r_L=0$ or $r_R=0$, with simultaneous satisfaction of unity transmittance $T=\vert t\vert^2=1$ for either cases.
The corresponding parametrization requires $\alpha = 0$ ($\beta =0$) for $r_R = 0$ ($r_L =0$) respectively.
However, Ref. \cite{yidong2} also indicates that certain accident situation can be occurred when both $\alpha =0$ and $\beta =0$ are met, a kind of bi-reflectionless.
Moreover, the transmission phase $\phi$ can be non-zero.
This result can be even observed at conventional ATR.
Due to the independence of $\phi$ involving CPAL and ATR cases, we can further plot 2D parametric space ($\alpha$ and $\beta$) to indicate extraordinary PT phenomena, as shown in Fig. 1 (c).
Green line represents CPAL, while red-dashed and blue-dashed lines represent ATR.
We can see there has a point at $\alpha=0$ and $\beta=0$, intersected by red-dashed and blue-dashed lines, corresponding to bi-reflectionless.

In symmetry phase, the eigenvalues of scattering matrix are unimodular in magnitudes and form complex conjugate pairs.
The parametric representation of symmetry phases is $\alpha\beta <0$, resulting in $T<1$.
We depict this symmetry phase by gray color in Fig. 1 (c).
For broken symmetry, the corresponding eigenvalues form reciprocal pairs in magnitudes.
This situation would be $\alpha\beta >0$, leading to  $T>1$.
We also depict this broken phase by bright yellow color as shown in Fig. 1 (c).
Thereby, the discrimination of symmetry and broken symmetry can be determined by the measurement of transmittance by simple single beam illumination.
In between bright yellow and gray region, it stands for exceptional points,  the degeneracy of eigenvectors, $\alpha=0$ or $\beta=0$, corresponding to reflectionless and unity transmittance,  i.e., $R_L=0$ or $R_R=0$, with $T=1$ both.

Furthermore, based on parametrization, we can display transmittance $T$, left reflectance $R_L$, and  right reflectance $R_R$ shown in Figs. 1 (d)-(f).
In Fig. 1 (d), we can obviously observe that $T<1$ denotes symmetry phase and $T>1$ denotes broken phase, while $T=0$ is expectational point.
Close to CPAL, $T$ would dramatically increase, while in CPA $T=\infty$.
In Figs. 1 (e) and (f), we observe the anisotropic property of reflectances from left and right illumination.
Interestingly, although PT systems definitely violate parity alone, In Figs. 1 (e) and (f), we find that the parity symmetry of reflection coefficients (PSR) for right and left is possible, i.e., $r_L=r_R$, when $\alpha=-\beta$.
We mark this PSR by a black dashed line.
PSR can occur not only  at symmetry phase but also at exceptional point.
For the latter, it is only at $\alpha=0$ and $\beta=0$,  corresponding to bi-reflectionless\cite{yidong2}.

\section{PT systems with extraordinary scattering events}
We investigate a PT heterostructure made of different composition for gain-gap-loss, shown in Fig. 2 (a).
Here are two composition in gain-loss: $\{G_1,L_1\}$ and $\{G_2,L_2\}$.
For each composition, we consider the following gain permittivity with a Lorentzian profile
\begin{equation}
\epsilon_{gain}(\omega)=\epsilon_0-\frac{\omega_{p,i}^2}{\omega_{0,i}^2-\omega^2+i\omega\gamma_i}
\end{equation}
here $i=1,2$ denotes different composition, $\omega_0$ is angular resonance frequency, $\omega_p$ is angular plasma frequency, and $\gamma$ is linewidth.
We let $\epsilon_0=1$ in the following calculation.
This permittivity expression satisfies Kramers-Kronig relation and causality \cite{kkrelation}.
For group $\{G_1,L_1\}$, the material parameters are $\omega_{p,1}^2=5\omega_0^2$  $\omega_{0,1}^2=\omega_0^2$, and $\gamma_{1}=0.3\omega_0$, while for another one $\{G_2,L_2\}$, $\omega_{p,2}^2=4\omega_0^2$,  $\omega_{0,2}^2=1.5\omega_0^2$, and $\gamma_{2}=1.2\omega_0$.
The geometry for this heterostructure system is by $l_1=0.431$ and $l_p=0.506$.
Now, we tune the operating frequency $\omega$ from $1.2\omega_0$ to $1.5\omega_0$.
In Fig. 2 (b), we show the corresponding parametric evolution by this operating window.
When $\omega=[1.21\omega_0,1.245\omega_0,1.35\omega_0]$, the system can exhibit PSR, ATR (from left), and CPAL (close).
By exploiting 2D parametric space, it not only provides the complete information in parameters but also demonstrate the corresponding transmittance, reflectances, and various PT phases.
Moreover, we can also display the complete parametrization $\alpha$, $\beta$, and $\phi$ by this operating wavelength in Fig. 2 (c).
When $1.21\omega_0\leq\omega< 1.245\omega_0$, the system belongs to symmetry phase and $T<1$.
At $\omega=1.245\omega_0$, this system is operated at ATR, associated with degenerate scattering eigenvectors, $T=1$, and $R_L=0$.
Beyond that, i.e., $1.245\omega_0<\omega$, the system is at broken phase and $T>1$.

When operating frequency is close to $1.35\omega_0$, the system is at CPAL.
We can observe that  by tuning $\omega$ close to $1.35\omega_0$, the trajectory  touches the allowed-forbidden boundary in Fig. 2 (c).
When $\omega=[1.21\omega_0,1.245\omega_0,1.35\omega_0]$, the transmission phases are $\phi=[0.27,0.31,-0.40]$, respectively.
In principle, extraordinary PT system can possess various transmission phases.
We note that in $\omega=1.245\omega_0$, the parameters $[\alpha,\beta]=[-1,0]$, corresponding to $R_R=0$ and $T=1$ but $R_L\neq 0$, i.e., ATR.
As indicated in Figs. 1 (e) and (f), system operating at ATR can have a variety of $R_R$ or $R_L$.

Next, we discuss the spectra of transmittance and  reflectances in Fig. 2 (d).
In PSR, two reflectances are identical, in ATR transmittance is unity and $R_L=1$, and in CPAL $T=19.55$, $R_L=33.07$, and $R_R=10.41$.
These extraordinary PT phenomena are marked by red star, blue cross, and black star, respectively.
We also perform full-wave simulations by COMSOL Multiphysics in Figs. 2 (e)-(g), where we show the instant strength of electric field.
In (e), this result is PSR, that the electric field displays parity symmetry outside the PT system, but it has no such parity inside the system.
In (f), it is ATR, that the input wave from left would experience reflectionless, while the transmittance is simultaneously unity.
We also note that as already proved by Ref. \cite{yidong2}, the field inside PT system has parity symmetry.
However, when the input wave comes from right, it would encounter reflection, that the corresponding field distribution in opposite sides are different.
In (g), this result is close to CPAL, where all waves involving transmission and reflection are amplified.
In ideal CPAL, transmittance and two reflectances  should be all infinity.

\begin{figure*}
\centering
\includegraphics[width=18cm]{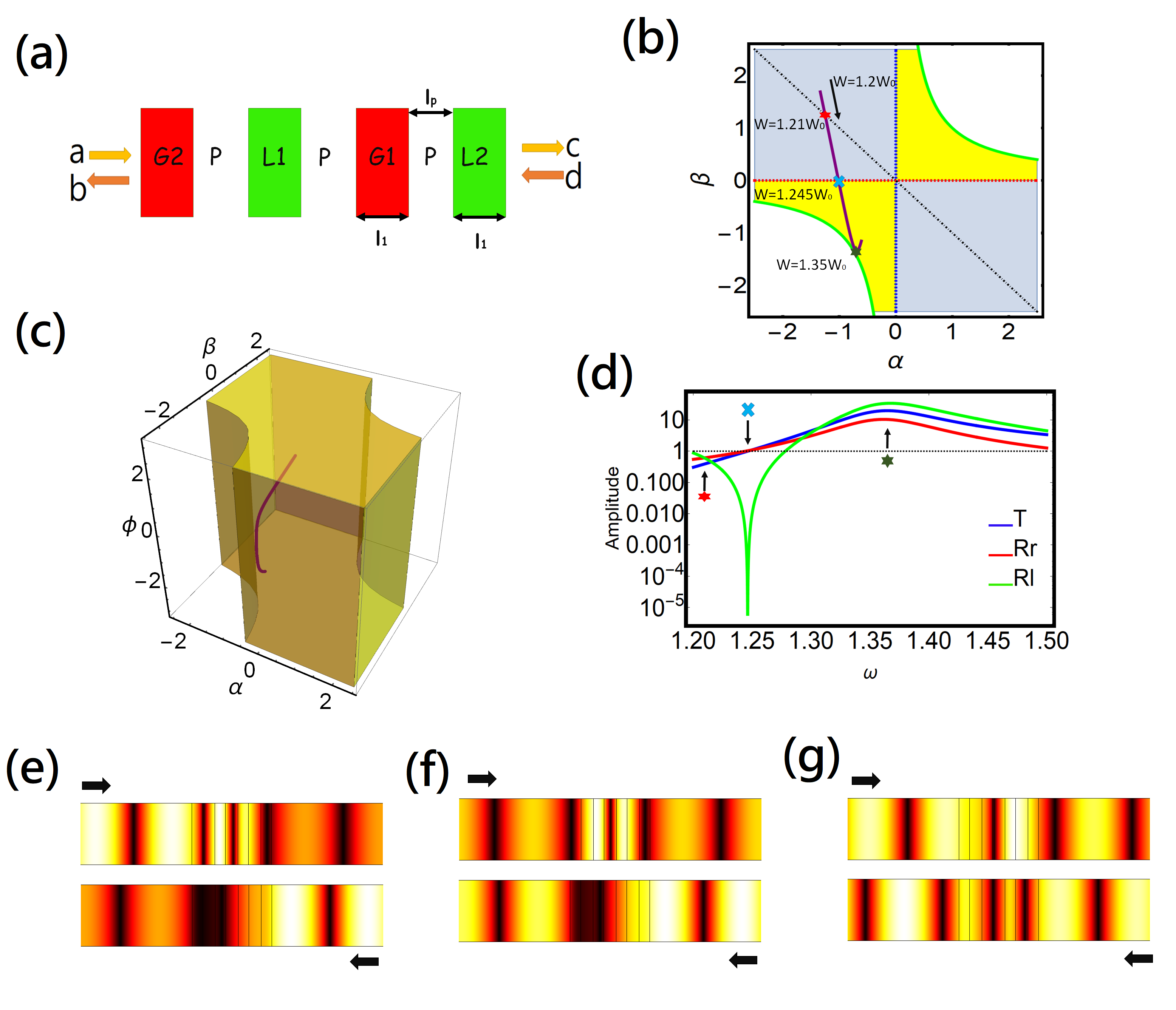}
\caption{(Color online) (a) Schematic configuration of a PT system made of two composition of gain-gap-loss. Each composition has its permittivity, Eq. (4), satisfied by Kramers-Kronig relation and causality. (b) By adjusting operating frequency from $\omega=[1.2\omega_0, 1.5\omega_0]$, we investigate its parametric variation of PT transfer matrix in (b) 2D ($\alpha$ and $\beta$) and (c) 3D ($\alpha$, $\beta$, and $\phi$). In the beginning $\omega=1.2\omega_0$, the PT belongs to symmetry phase regime. When $\omega=1.21\omega_0$, the PT has PSR situation.  When $\omega=1.245\omega_0$, it is ATR, associated with exceptional point (the emergence of  degenerate scattering eigenvectors). When $\omega= 1.35\omega_0$, the system response is close to CPAL and  belongs to broken phase.  In (c), we demonstrate a clear relation between transmission phase and these extraordinary scattering phenomena. In principle, transmission phases can be arbitrary. The corresponding transmittance and reflectance from right and left illumination are shown in (d). We also mark PSR, ATR, and CPAL, by red star, light blue cross, and black star, respectively. Comparison with the results provided by parametric spaces (b) and (c), the spectra provides a specific information related to system configuration. With COMSOL Multiphysics (a finite element
solver), we plot the corresponding strength of electric field in (e), (f), and (g) for PSR, ATR, and CPAL (close), respectively. As expected, the field in (e) has parity symmetry including inside and outside PT system. In (f), the input field is reflectionless from left side and the (absolute) amplitude of transmitted field is same as incident one, while the corresponding transmission phase is $0.31$.  As incident field from right side, there has a reflected wave, indicated in bottom side of (f). We note that in this opposite illumination, the field inside PT system would no long form a parity symmetry. In (g), this system performs CPAL (close). Enhanced transmitted and reflected waves from opposite illuminations are shown. In ideal CPAL, the amplitudes of transmitted and reflected waves are infinite, corresponding to amplifying.} 
\end{figure*}

\section{Design of functional PT system with  by parametrization}
Contributed by successful development of metasurface,  gain metasurface can be implemented by monolayer or bilayer graphene with population inversion \cite{graphene1,graphene2}.
Definitely, it is a long-sought goal to have compact PT photonic systems.
Now, we turn to consider another  subwavelength PT system made of a ultra-thin gain (denoted as G) and a ultra-thin loss (denoted as L) separated by a gap (denoted as P), as shown in Fig. 3 (a).
Our interest here is to manipulate this simple PT system having various PT functionality with proper material parameters and geometrical sizes with the help of  parametrization.

Due to that its physical dimension of gain and loss slabs is much smaller than operating wavelength, thus we can approximate $\sin[nk_1l_1]\approx n k_1l_1$ and $\cos[n k_1l_1]\approx 1$ to the first order,  here $n$ is index of refraction . 
Furthermore, we can obtain one approximated transfer matrix.
The detail of derivation is placed at Appendix A.
Now, based on these expressions, the corresponding material permittivities for loss are found,
 \begin{widetext} 
\begin{equation}
\begin{split}
\epsilon_l^{''}(\alpha,\beta,l_1,l_2)&=\frac{\alpha+\beta}{k_0l_1[2\sin(k_0l_2)+2k_0l_1\cos(k_0l_2)]}\\
\epsilon_l^{'\pm}(\alpha,\beta,l_1,l_2)&=\frac{\cos[k_0l_2]\pm \sqrt{\cos^2[k_0l_2]-\sin[k_0l_2][\alpha-\beta+2k_0l_1\cos[k_0l_2]+(k_0l_1)^2(-1+(\epsilon_l^{''})^2)\sin[k_0l_2]]}}{k_0l_1\sin[k_0l_2]}.
\end{split}
\end{equation}
\end{widetext}
where $l_1$ is thickness of gain/loss slab, $l_2$ is distance of gap, $\epsilon_l^{'\pm}$ and $\epsilon_l^{''}$ denote the real part and  imaginary part of loss permittivity, respectively.
We note that there are two solutions for the real part of loss permittivity, $\epsilon_l^{'\pm}$.
In the following calculation, we set $k_0=1$.
Obviously, the values of $\epsilon_l^{'\pm}$ and $\epsilon_l^{''}$ depend on not only parametrization $\alpha$ and $\beta$, but also geometry parameters $l_1$ and $l_2$.
Thus, under $\alpha$ and $\beta$ given, by tuning $l_1$ or $l_2$, we can find out the corresponding material permittivities to support our required functionality.

Once we know the material permittivities, we can further calculate the transmission phase, $\phi$,
 \begin{widetext} 
\begin{equation}
\phi=\frac{1}{i}\ln \{\frac{1}{\sqrt{1-\alpha\beta}}[\frac{1}{4}e^{-ik_1l_2}k_1^2l_1^2(-1+\epsilon_l^{'}-i\epsilon_l^{''})(-1+\epsilon_l^{'}+i\epsilon_l^{''})+e^{ik_1l_2}[1+\frac{1}{2}ik_1l_1(1+\epsilon_l^{'}-i\epsilon_l^{''})][1+\frac{1}{2}ik_1l_1(1+\epsilon_l^{'}+i\epsilon_l^{''})]]\}.
\end{equation}
\end{widetext}

To verify our strategy, we choose two different sets of parametrization $\alpha$ and $\beta$ to figure out the corresponding material permittivities for gain and loss, marked by red and blue dots in Fig. 3 (b).
For red dot, the parametrization $[\alpha,\beta]=[0.4,0.4]$, belonging to broken phase regime and $T>1$.
Next, we first give the distance of gap by $l_2=2.24$, then we estimate the corresponding loss permittivity in Fig. 3 (c).
Here data marked by blue points come from the results of analytical expressions, Eq.(5), which coincide with numerical calculation by orange points.

In another set of parametrization $[\alpha,\beta]=[-2,0]$, it denotes ATR, belonging to exceptional point regime, $T=1$ and $R_L=0$.
Here we choose $l_2=1.45$ as a given parameter, then we calculate the corresponding loss permittivity in Fig. 3 (d).
Again, these data obtained from analytical expression and numerical calculation can match so well.
As a result, through tuning geometry parameters, one could discover the realizable material parameters in order to satisfy the required PT functionality at will.
Finally, we note that the concept of PT parametric space can be applied to one-dimensional PT-symmetric photonic networks with any arbitrary configurations, even for metasurfaces and those constituted by multiple spatially-distributed, balanced gain and loss elements. 

\section{Conclusion}
We herein introduce a generalized parametric space for any PT systems, with arbitrary system configurations.
The resulting parametric space, which stems from the nature of PT symmetry and non-reciprocity, can provide the information needed for realization of novel devices.
Parametric space can provide all information for various PT phases, transmittance, reflectances, and extraordinary wave phenomena including CPAL, ATR, and PSR.
Through the parametric diagram, we find that PT system can have parity symmetry of reflection, even although PT system violates parity alone.
We demonstrate a PT heterostructure to support this finding.
Furthermore, we also study a subwavelength PT system with ultra thin composition in gain and loss.
We further obtain the analytical material expressions for loss and gain, depending on parametrization.
In such a manner, the design of novel PT devices becomes controllable and flexible.
Our results could provide a clear guideline for design of optical and photonic PT devices and could be readily extended to the realm of acoustics, elastics and other wave physics.

\begin{figure*}
\centering
\includegraphics[width=18cm]{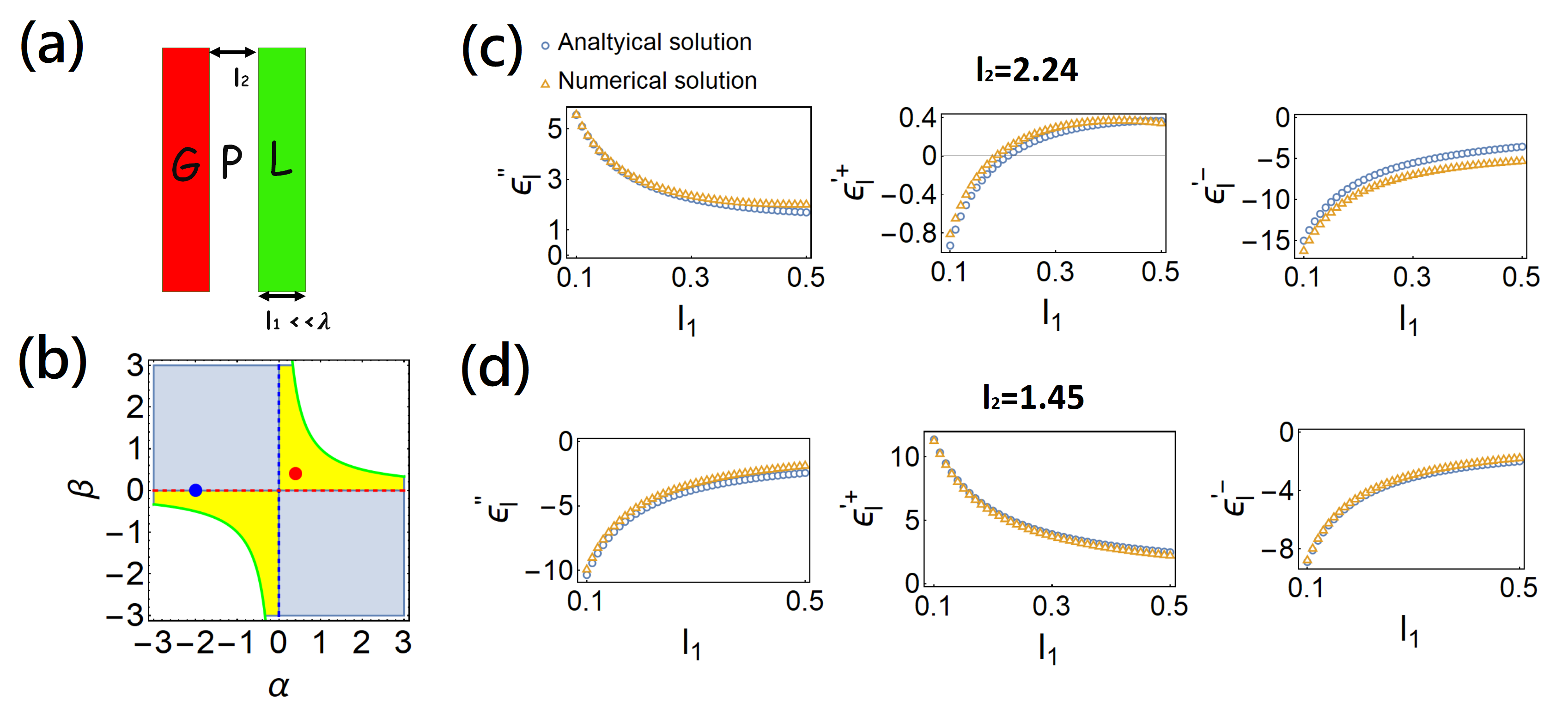}
\caption{(Color online) (a) Schematic configuration of a PT system made of a ultra-thin gain (denoted as G) and a ultra-thin loss (denoted as L) separated by a gap (denoted as P). We choose two sets of parametrization to manipulate PT phenomena marked by red and blue dots in (b). Under $l_2$ given, we can find  the corresponding lossy permittivity of Eq. (5) by tuning $l_1$ from $0.1$ to $0.5$. Orange points denote the results by numerical calculation, while blue ones denote the results from analytical expression, Eq. (5). Numerical and analytical results can match so well. } 
\end{figure*}

\section{Acknowledgement}

This work was supported by Ministry of Science and Technology, Taiwan (MOST) ($107$-
$2112$-M-$143$-$001$-MY3).

\section*{Appendix A} 

We discuss a PT slab system made by a gain (denoted as G), a gap (denoted as P) and a loss (denoted as L) as shown in Fig. 3 (a).
The resultant transfer matrix is
\begin{equation}
\begin{split}
M_{tot}=M_{slab}(n_lk_1,l_1)M_t(l_2)M_{slab}(n_gk_1,l_1)
\end{split}
\end{equation}
here $n_l^2=\epsilon_l^{'}+i\epsilon_l^{''}$ ($n_g^2=n_l^{*2}=\epsilon_l^{'}-i\epsilon_l^{''}$) is lossy (gain) index of refraction, $l_1$ is the thickness of the slab, and $l_2$ is gap distance.
The slab transfer matrix, \cite{book1}, is 
 \begin{widetext} 
\begin{equation}
\begin{split}
M_{slab}(k_2,l_2)=
\begin{bmatrix}
\cos[k_2l_2]+\frac{i}{2}\sin[k_2 l_2][\frac{k_1}{k_2}+\frac{k_2}{k_1}] & \frac{i}{2}\sin[k_2l_2][\frac{k_2}{k_1}-\frac{k_1}{k_2}]\\
-\frac{i}{2}\sin[k_2l_2][\frac{k_2}{k_1}-\frac{k_1}{k_2}] & \cos[k_2l_2]-\frac{i}{2}\sin[k_2 l_2][\frac{k_1}{k_2}+\frac{k_2}{k_1}] 
\end{bmatrix}
\end{split}
\end{equation}
 \end{widetext} 
here the wavenumber in the slab is $k_2$ and the thickness of slab is $l_2$.
The transfer matrix $M_t(l_2)$ is
\begin{equation}
\begin{split}
M_t(l_2)=
\begin{bmatrix}
e^{ik_2l_2} & 0\\
0 & e^{-ik_2l_2}
\end{bmatrix}.
\end{split}
\end{equation}

By using small approximation to the first order expansion, $\sin[kl]\approx kl$ and $\cos[kl]\approx 1$, we can approximate $M_{tot}\rightarrow M_{tot}^{app}$,
 \begin{widetext} 
\begin{equation}
\begin{split}
M^{app}_{tot,11}&=\frac{1}{4}e^{-ik_1l_2}k_1^2l_1^2(-1+n_g^2)(-1+n_l^2)+e^{ik_1l_2}[1+\frac{1}{2}ik_1l_1(1+n_g^2)][1+\frac{1}{2}ik_1l_1(1+n_l^2)]\\
M^{app}_{tot,12}&=\frac{1}{4}e^{-ik_1l_2}k_1l_1[(2i+k_1l_1(1+n_g^2))(-1+n_l^2)-e^{2ik_1l_2}(-1+n_g^2)(-2i+k_1l_1(1+n_l^2))]\\
M^{app}_{tot,21}&=\frac{1}{4}e^{-ik_1l_2}k_1l_1[e^{2ik_1l_2}(-2i+k_1l_1(1+n_g^2))(-1+n_l^2)-(-1+n_g^2)(2i+k_1l_1(1+n_l^2))]\\
M^{app}_{tot,22}&=\frac{1}{4}e^{-ik_1l_2}[4-2ik_1l_1(2+n_g^2+n_l^2)+k_1^2l_1^2(e^{2ik_1l_2}(-1+n_g^2)(-1+n_l^2)-(1+n_g^2)(1+n_l^2))].\\
\end{split}
\end{equation}
 \end{widetext} 

Then, solving $M^{app}_{tot,12}+M^{app}_{tot,21}=i(\alpha+\beta)$ and $M^{app}_{tot,12}-M^{app}_{tot,21}=i(\alpha-\beta)$, we obtain
\begin{equation}
\begin{split}
\epsilon_l^{''}&=\frac{\alpha+\beta}{k_1l_1[2\sin(k_1l_2)+2k_1l_1\cos(k_1l_2)]}\\
\epsilon_l^{'}&=\frac{\cos[k_1l_2]\pm \sqrt{\cos^2[k_1l_2]-\sin[k_1l_2][\alpha-\beta+2k_1l_1\cos[k_1l_2]+(k_1l_1)^2(-1+(\epsilon_l^{''})^2)\sin[k_1l_2]]}}{k_1l_1\sin[k_1l_2]}
\end{split}
\end{equation}

\pagebreak
The transmission phase could be found by inversely solving $M^{app}_{tot,11}=\sqrt{1-\alpha\beta}e^{i\phi}$,
 \begin{widetext} 
\begin{equation}
\phi=\frac{1}{i}\ln \{\frac{1}{\sqrt{1-\alpha\beta}}[\frac{1}{4}e^{-ik_1l_2}k_1^2l_1^2(-1+\epsilon_l^{'}-i\epsilon_l^{''})(-1+\epsilon_l^{'}+i\epsilon_l^{''})+e^{ik_1l_2}[1+\frac{1}{2}ik_1l_1(1+\epsilon_l^{'}-i\epsilon_l^{''})][1+\frac{1}{2}ik_1l_1(1+\epsilon_l^{'}+i\epsilon_l^{''})]]\}.
\end{equation}
\end{widetext} 


\begin{thebibliography}{99}
\bibitem{note}
There has another description for scattering matrix $S_{1}$ by $S_1=S\sigma_1$ where $\sigma_1=\sigma_x$ is Pauli matrix.
Two expressions would result in 
discrepancy results for eigenvalues, leading to different determination of symmetry-symmetry breaking.
In this context, we adopt the $S$, directly connected with transmittance measurement by only a single beam detection.
\end{thebibliography}
\end{document}